\documentclass[pra,twocolumn,numerical,floatfix,superscriptaddress]{revtex4-1}

\newcommand{\ket}[1]{| #1 \rangle}
\newcommand{\bra}[1]{\langle #1 |}

\usepackage[english]{babel}
\usepackage{amsmath,amssymb}
\usepackage{acronym}
\usepackage{setspace}

\usepackage{graphicx}

\usepackage{overpic}
\usepackage{rotating}
\usepackage{subfigure}
\usepackage{multirow}

\begin{document}

\title{Low-Loss  All-Optical Zeno Switch in a Microdisk Cavity Using EIT}

\author{B. D. Clader}
\email{dave.clader@jhuapl.edu}
\affiliation{The Johns Hopkins University Applied Physics Laboratory, Laurel, MD, USA}
\author{S. M. Hendrickson}
\affiliation{The Johns Hopkins University Applied Physics Laboratory, Laurel, MD, USA}
\author{R. M. Camacho}
\affiliation{Sandia National Laboratories, Albuquerque, NM, USA}
\author{B. C. Jacobs}
\affiliation{The Johns Hopkins University Applied Physics Laboratory, Laurel, MD, USA}

\begin{abstract}We present theoretical results of a low-loss all-optical switch based on \acl{EIT} and the classical  Zeno effect in a microdisk resonator. We show that a  control beam can modify the atomic absorption of the evanescent field which suppresses the cavity field buildup and alters the path of a weak signal beam.  We predict more than 35 dB of switching contrast with less than 0.1 dB loss using just 2 $\mu$W of control-beam power for signal beams with less than single photon intensities inside the cavity.
\end{abstract}

\maketitle

\acrodef{SiN}{Silicon Nitride}
\acrodef{Cs}{Cesium}
\acrodef{Rb}{Rubidium}
\acrodef{TPA}{two-photon absorption}
\acrodef{SPA}{single photon absorption}
\acrodef{SEM}{scanning electron microscope}
\acrodef{cavRes}[$\lambda_{c}$]{cavity resonance}
\acrodef{780Res}[$\lambda_{Rb}^{780}$]{780 atomic resonance}
\acrodef{1529Res}[$\lambda_{Rb}^{1529}$]{1529 atomic resonance}
\acrodef{HF}{Hydrofluoric acid}
\acrodef{Q}{quality factor}
\acrodef{QZE}{quantum Zeno effect}
\acrodef{AZE}{anti Zeno effect}
\acrodef{FOM}{figure of merit}
\acrodef{EIT}{electromagnetically induced transparency}
\acrodef{AOM}{acousto-optic modulator}
\acrodef{EOM}{electro-optic modulator}
\acrodef{SERF}{spin-exchange-relaxation-free}
\acrodef{SQUID}{superconducting quantum interference device}
\acrodef{SERS}{stimulated electronic Raman scattering}
\acrodef{PMT}{photo-multiplier tube}

%
%
%
%
\section{Introduction}
Over the past few decades transistors and other computing components have dropped in size while simultaneously increasing performance.  However, power dissipation is increasingly becoming a fundamental limitation to performance \cite{1250885}.   All-optical switches and transistors seek to address this issue, while simultaneously pushing forward the technology needed to create all-optical quantum computing devices \cite{miller2010optical, Dawes29042005, hu2008picosecond, Waldow:08, albert2011cavity}.  One of the fundamental issues limiting this technology is the strength of the nonlinear effects that couple the signal and control fields.  Electromagnetically induced transparency (\acsu{EIT})  \cite{harris:36, RevModPhys.77.633} has been investigated as a resource for optical switches \cite{PhysRevLett.102.203902, Zhang:07, Fleischhauer02092011, PhysRevA.71.043811} and quantum memories \cite{PhysRevLett.86.783, PhysRevLett.88.023602, PhysRevA.65.031802, RevModPhys.75.457} due to its large nonlinearity, which is enhanced by coherent effects. We demonstrate how it can be used along with a microdisk resonator to create a high-speed low-loss all-optical switch.

The \ac{QZE} \cite{misra1977zeno} is a process whereby frequent measurements of a quantum system can inhibit transitions.  In the first experimental demonstration, the \ac{QZE} was shown to inhibit driven transitions between ${}^9{\rm Be}^+$ ground-state hyperfine levels \cite{PhysRevA.41.2295}.   It has been demonstrated that a classical analogue to the \ac{QZE} can be used to create an all-optical switch using \ac{TPA} in a resonant optical cavity \cite{jacobs2009all, ZenoSwitch}.  This switch consists of a four-port resonator evanescently coupled to \ac{Rb} vapor.  The presence of two input beams results in sufficiently strong \ac{TPA} to suppress the cavity field build-up thereby altering the path of the beams due to inteference effects.  All-optical Zeno switches have also been proposed using other nonlinearities including Raman  \cite{Wen:11} and inverse Raman induced loss \cite{Kieu:12}.

In addition to classical switching applications, it was shown that the Zeno effect could be a used for quantum information purposes.  With a sufficiently strong nonlinearity, it could enable quantum logic gates \cite{knill2001scheme, kok2007linear} without the need for a large number of ancilla photons.  The ability of the \ac{QZE} to inhibit transitions could be used instead of protective ancilla photons to ensure that error events are suppressed in photonic quantum gates \cite{PhysRevLett.85.1762, PhysRevA.70.062302, Franson:07, PhysRevA.77.062332, Shao:09}.  Related work has studied the \ac{QZE} as a tool for protecting entanglement \cite{PhysRevLett.100.090503} and its effect on entanglement \cite{PhysRevA.82.052118} in general.  It has also been suggested that the \ac{QZE} could be used to implement quantum logic gates using semiconductor materials \cite{PhysRevLett.103.037401}.

Here we report theoretical results and performance estimates showing that \ac{EIT} and Autler--Townes splitting \cite{PhysRev.100.703} can also be used to implement an all-optical switch in a microdisk cavity based on the Zeno effect.  We make use of \ac{SPA} to suppress the resonant field buildup in a cavity, and use a control beam to modulate the absorption by inducing \ac{EIT} (we refer to the combined \ac{EIT} and Autler--Townes splitting effect as simply \ac{EIT} from here out).  The benefit of this approach is that on-resonant \ac{SPA} has a higher absorption cross-section than known nonlinear processes, potentially enabling better switching results as long as \ac{SPA} can be sufficiently reduced by \ac{EIT}.  In addition this approach allows the use of very weak signal beams, making it a good candidate to switch single-photon intensities with low loss.  

\begin{figure}
\begin{center}
\subfigure[\hspace{1pt}EIT Off -- SPA On]{
\hspace{9pt}
\begin{overpic}[width=6.0cm]{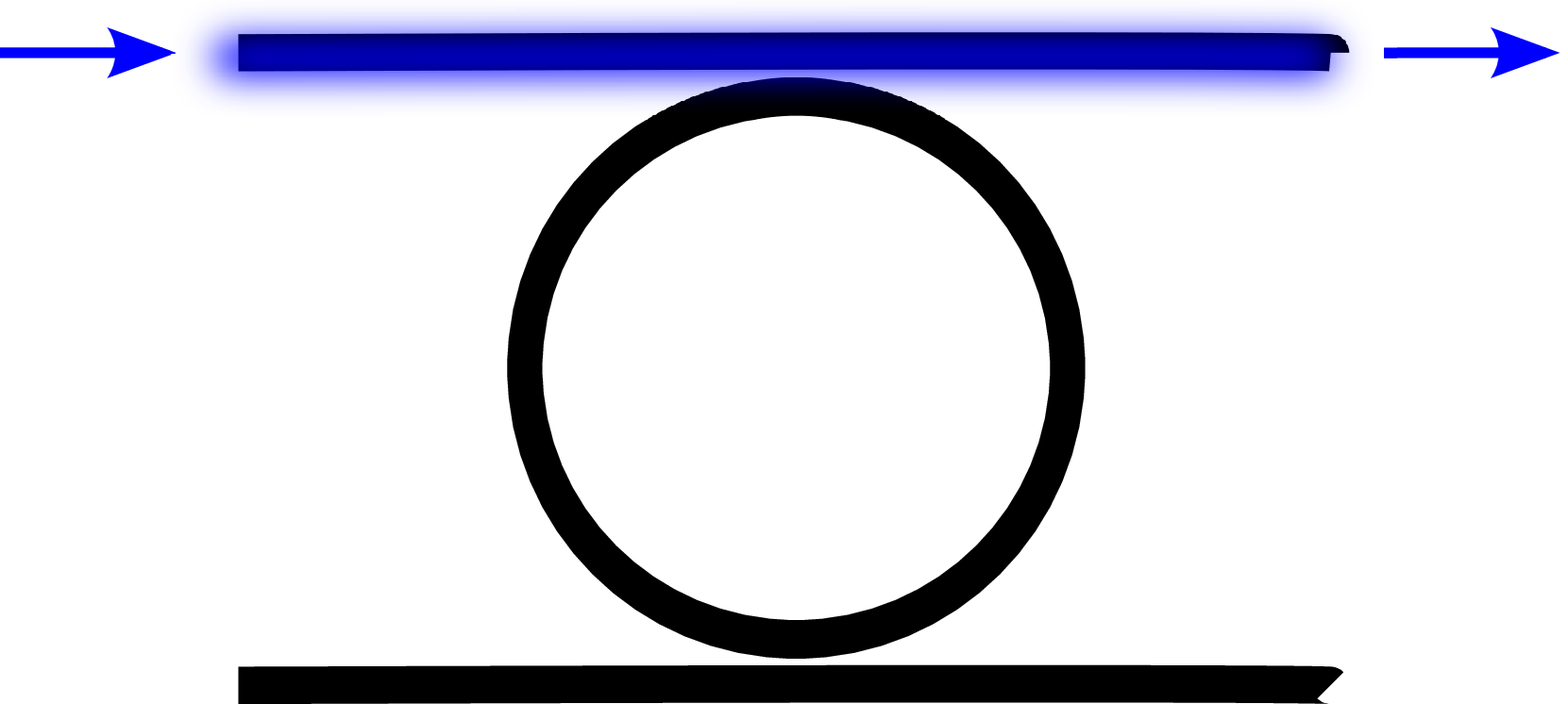}
\put(46,20){SPA}
\put(1,46){$E_{in}$}
\put(90,46){$E_{t}$}
\end{overpic}
\vspace{2pt}
\label{fig:resonator_spa}
}
\hspace{2cm}
\subfigure[\hspace{1pt}EIT On]{
\hspace{-20pt}
\begin{overpic}[width=5.3cm]{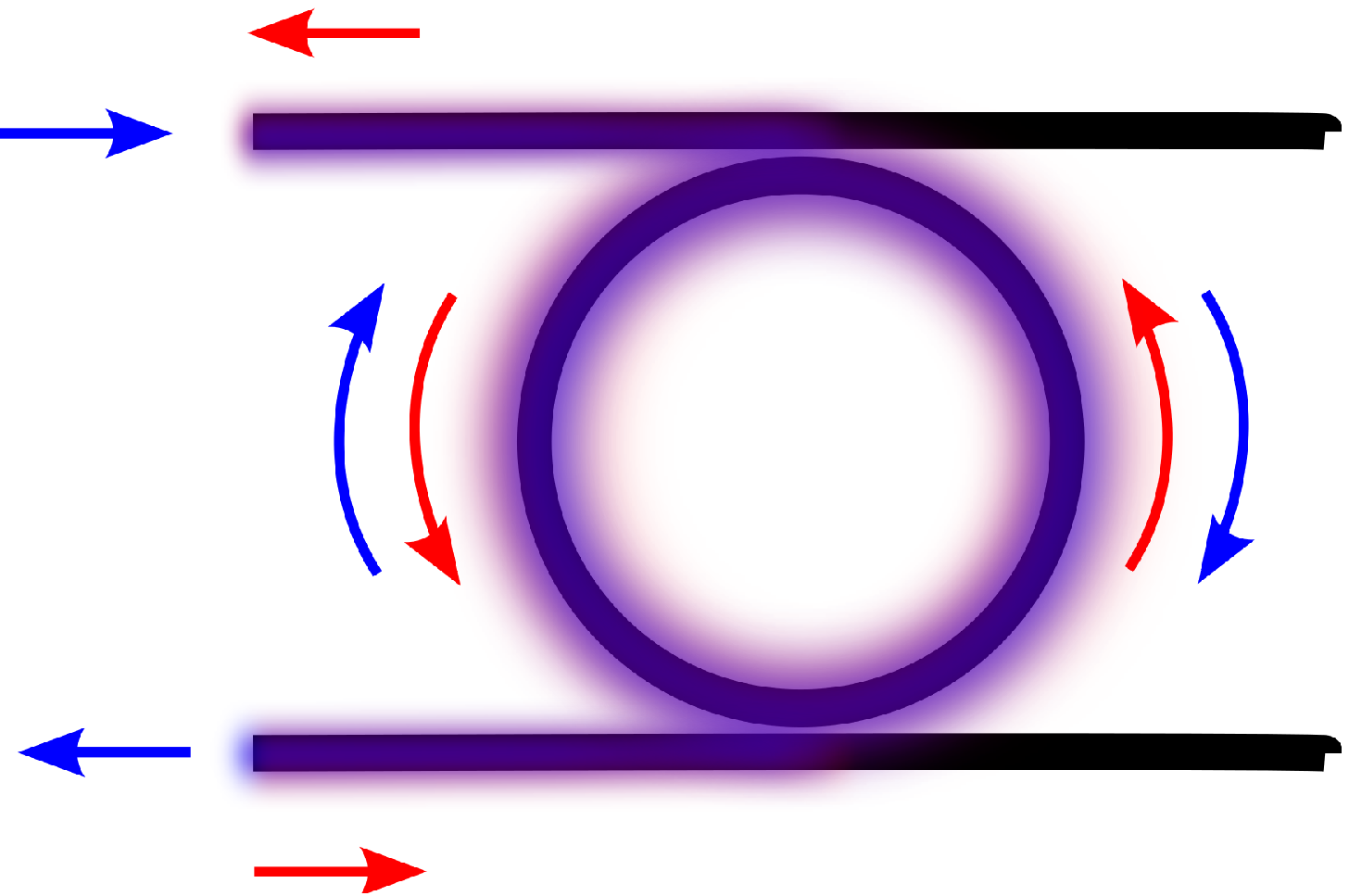}
\put(54,32){EIT}
\put(2,62){$E_{in}$}
\put(2,16){$E_{d}$}
\end{overpic}
\label{fig:resonator_eit}
}
\end{center}
\caption{\label{fig:eit_switch}Schematic of the EIT based Zeno all-optical micro-resonator switch.
\\
(a) When only a single input beam is weakly coupled to the cavity, strong \ac{SPA} inhibits field buildup causing the beam to bypass the resonator and exit via the through port $E_t$.
\\
(b) When two beams are present, denoted by blue for the signal beam and red for the control beam, the control beam eliminates the evanescent coupling of the fields to the atoms surrounding the cavity through \ac{EIT}.  This reduces the loss present, allowing the signal beam to build in the resonator, exiting through the drop port $E_d$.}
\end{figure}

To understand how the switch operates, consider the operation of a four-port resonator, as shown in Fig. \ref{fig:eit_switch}, in the absence of any atomic interaction (see e.g. \cite{haus1984} and \cite{vahala2004} for detailed discussions of four-port resonators and nomenclature).  On resonance, a weakly-coupled signal $E_{in}$ will be almost completely transferred to the drop-port as a result of destructive interference between the resonant build-up of the cavity field and the light in the input waveguide.  Next, consider how this system would change with the addition of a strong loss mechanism in the cavity, such as shown in Fig. \ref{fig:resonator_spa}.  The resonant field buildup of the signal beam will be suppressed which dramatically reduces the destructive interference in the input waveguide and results in nearly all of the incident light exiting the through-port.  Counter-intuitively, the presence of the strong loss mechanism will not dramatically increase the loss of the system but will instead alter the coupling condition of the cavity which changes the output path of the light.  

This is directly analogous to the suppression of probability amplitudes via measurement in the quantum Zeno effect.  This is made more clear when considering the case of single photons in the signal beam.  In that case, a null absorption (measurement) event by the atoms surrounding the cavity causes the photon to bypass the resonator.  The stronger this potential absorption process, the more likely that a null absorption event occurs and the photon bypasses the cavity.  More information on this can be found in Refs. \cite{PhysRevA.70.062302, jacobs2009all}.

Now we consider a loss mechanism based on a cascade \ac{EIT} scheme using the $5S_{1/2}$, $5P_{3/2}$ and $5D_{5/2}$ states of \ac{Rb}.  We take the signal beam to be resonant with the $5S_{1/2} \rightarrow 5P_{3/2}$ transition near 780 nm and the control beam to be resonant with the $5P_{3/2} \rightarrow 5D_{5/2}$ transition near 776 nm.  When the 776 nm \ac{EIT} control beam is present, a transmission window is created in the 780 nm single-photon absorption line.  This allows the signal beam to build in the resonator and exit through the drop port, as shown schematically in Fig. \ref{fig:resonator_eit}.  

In this paper, we theoretically analyze the performance of such a device using high-fidelity numerical models and present detailed estimates of switching performance.  Our results indicate that this scheme enables high-contrast, low-loss switching at timescales on the order of the total cavity relaxation time which is roughly $\sim 100$ picoseconds for the devices under consideration in this paper.  

%
%
%
%
\section{Theoretical Model}

We model the transmission characteristics of the four-port microdisk shown in Fig. \ref{fig:eit_switch} evanescently coupled to Rubidium vapor. The device specifications have been chosen to be consistent with current fabrication capabilities.  The design consists of an unclad, suspended Si$_3$N$_4$ disk with a free spectral range equal to the 4 nm difference between the $5S_{1/2} \to 5P_{3/2}$ $D_2$ line at 780 nm and the $5P_{3/2} \to 5D_{5/2}$ line at 776 nm, allowing for simultaneous resonance at 776 nm and 780 nm (for an example of a similar device operating at a higher wavelength see Ref. \cite{ZenoSwitch}). The thickness is chosen such that roughly 30\% of the electric field energy of the fundamental mode is outside the resonator, allowing considerable interaction with the Rubidium vapor surrounding the cavity.

\begin{figure}
\begin{center}
\begin{overpic}[width = 8.0cm]{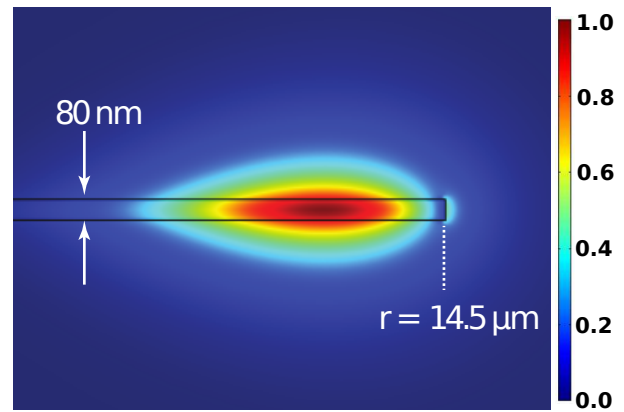}\end{overpic}
\caption{\label{fig:FieldProfile}Simulated normalized electric field profile of the microdisk, showing the radial component of the mode used in the quantum calculation.  The cutout is the along page (side) view of Fig. \ref{fig:eit_switch}.  The radius of the device is set such that the free spectral range is equal to 4 nm, allowing for simultaneous resonance at 780nm and 776nm, and the thickness of the disk is chosen such that 30\% of the field is outside the resonator.}
\end{center}
\end{figure}

We estimate the field profile of the fundamental mode in the cavity using a fully vectorial 2-D axially symmetric weighted residual formulation of Maxwell's equations implemented in Comsol Multiphysics software (as shown in Fig. \ref{fig:FieldProfile}).  The field profile is used to estimate the evanescent interaction of signal and control beams with an ensemble of three-level atoms by calculating the effective absorption coefficient.  This calculated absorption coefficient is combined with a classical model of a resonator to predict switching performance. The intent is to show that the presence of the control beam will induce \ac{EIT}, modifying the absorption coefficient of the signal beam in the resonator.  This causes a change in the overall coupling between the waveguide and the resonator resulting in switching.

%
%
%
%
\subsection{Atomic Model}
\begin{figure}[h!]
\begin{center}
\begin{overpic}[height = 2.2in]{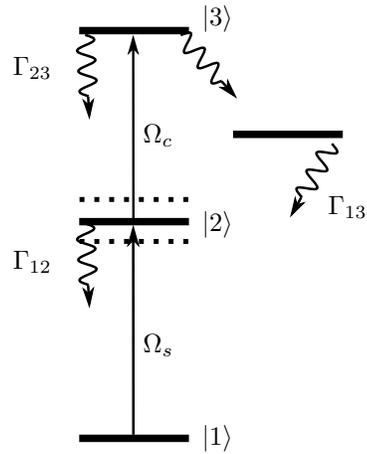}
\put(-15,42 ){$\Gamma_{12}$}
\put(-15,88){$\Gamma_{23}$}
\put(60,55){$\Gamma_{13}$}
\put(30,0){$\ket{1}$}
\put(30,51){$\ket{2}$}
\put(30,100){$\ket{3}$}
\put(16,22){$\Omega_s$}
\put(16,72){$\Omega_c$}
\end{overpic}
\caption{\label{fig:LadderSystem}The three-level atom we use to model the resonant $5S_{1/2} \to 5P_{3/2}$ and $5P_{3/2} \to 5D_{5/2}$ transitions in Rubidium, including all decay channels. The dotted lines indicate the $5P_{3/2}$ splitting induced by the upper beam.  The state on the right is the $6P_{3/2}$ which provides a decay channel from the $5D_{5/2}$ excited state to the $5S_{1/2}$ ground state.}
\end{center}
\end{figure}

The device under consideration is designed to support simultaneous cavity resonances at the 780 nm and 776 nm spectral lines of Rubdium.  To model this interaction between the cavity fields and the Rubidium vapor, we approximate the Rubidium atom as the four-level atomic system shown in Fig. \ref{fig:LadderSystem}.  The fourth level, indicated on the right, is the decay channel from the excited state level $\ket{3}$ to the ground state level $\ket{1}$ through the $6P_{3/2}$ level.  It is included in our atomic model only as a decay term.  The Hamiltonian of this system is the standard three-level cascade model given by
\begin{align}
\label{eq:hamiltonian}
H  &= \left(-\hbar\frac{\Omega_s}{2}\ket{1}\bra{2} - \hbar\frac{\Omega_c}{2}\ket{2}\bra{3} + \textnormal{h.c.}\right) \\ \nonumber
& + \hbar\Delta_s|1\rangle\langle 1| - \hbar\Delta_c|3\rangle\langle 3|.
\end{align}
We denote all variables associated with the 776 nm control field and 780 nm signal field with the subscripts c and s respectively.  The Rabi frequency is $\Omega=2 \vec{d}\cdot\hat{\epsilon}\mathcal{E}/\hbar$, where $\vec{d}$ is the dipole moment, $\mathcal{E}$ is the slowly varying electric field amplitude, $\hat{\epsilon}$ is the polarization vector, and $\Delta$ is the detuning of the field from its associated transition.  This Hamiltonian, along with the addition of phenomenological decay terms, gives the following density matrix equations:
\begin{subequations}
\label{eq:densitymatrix}
\begin{align}
\dot{\rho}_{11} &= i\frac{\Omega_s}{2}\rho_{21} - i\frac{\Omega_s^*}{2}\rho_{12} + \Gamma_{12}\rho_{22}+\Gamma_{13}\rho_{33} \\
\dot{\rho}_{22} &= i\frac{\Omega_c}{2}\rho_{32} - i\frac{\Omega_c^*}{2}\rho_{23} + i\frac{\Omega_s^*}{2}\rho_{12} - i\frac{\Omega_s}{2}\rho_{21} \\ \nonumber
& -\Gamma_{12}\rho_{22} + \Gamma_{23}\rho_{33} \\
\dot{\rho}_{33} &= i\frac{\Omega_c^*}{2}\rho_{23} - i\frac{\Omega_c}{2}\rho_{32}  - (\Gamma_{13} + \Gamma_{23})\rho_{33}\\
\dot{\rho}_{12} &= i\frac{\Omega_s}{2}\left( \rho_{22}-\rho_{11}\right) -i \frac{\Omega_c^*}{2}\rho_{13} - i(\Delta_s-i\gamma_{12})\rho_{12} \\
\dot{\rho}_{13} &= i\frac{\Omega_s}{2}\rho_{23} - i\frac{\Omega_c}{2}\rho_{12} - i(\Delta_s + \Delta_c - i\gamma_{13})\rho_{13} \\
\dot{\rho}_{23} &= i\frac{\Omega_c}{2}\left( \rho_{33} - \rho_{22}\right) -i\frac{\Omega_c}{2}\rho_{22} - i(\Delta_c - i\gamma_{23}) \rho_{23},
\end{align}
\end{subequations}
where $\Gamma_{mn}$ is the homogeneous decay term associated with the $m \to n$ transition, and $\gamma_{mn}$ is the transverse decay term associated with the off-diagonal elements.  We take $\gamma_{12} = \Gamma_{12}/2$, $\gamma_{13} = (\Gamma_{13}+\Gamma_{23})/2$,  and $\gamma_{23} = (\Gamma_{12} + \Gamma_{13} + \Gamma_{23})/2$.  These off-diagonal decay rates assume that all decoherence is due to population decay, which is suitable for atomic vapors.  The values were chosen to be physically consistent with a positive density matrix, which does not always hold for arbitrary decoherence terms \cite{PhysRevA.70.022107, PhysRevA.71.022501}.  The branching ratio between the $\Gamma_{23}$ and $\Gamma_{13}$ decay channels was set to $0.65:0.35$ as given in Ref. \cite{HEAVENS:61}.

Many analytic solutions to Eqs. \eqref{eq:densitymatrix} and similar models have been analyzed \cite{JETP.30.243, PhysRevA.11.1641, PhysRevA.51.576, RevModPhys.77.633}.   However, to ensure accurate device performance estimates, we use numerical solutions to the density matrix equations above.  This allows us to solve Eqs. \eqref{eq:densitymatrix}, with only a steady-state approximation, which is consistent with out experiments using CW lasers.  However we do not make any perturbation approximations since we assume a high-Q resonator, which can develop intense fields.   The steady-state approximation can easily be lifted to model pulsed phenomena, however we include it here in order to reduce the run-time of our numerical simulations.

We include Doppler broadening in our simulations to account for thermal motion of the atoms.  We do so by averaging the density matrix elements over the Doppler profile given by
\begin{equation}
\label{eq:doppler}
F(\Delta) = \frac{1}{\sqrt{2 \pi}\sigma_D}e^{-\frac{\Delta^2}{2\sigma_D^2}},
\end{equation}
where $\Delta$ is the laser field detuning due to the thermal velocity of the atom, and $\sigma_D$ is the Doppler linewidth, given by the Maxwell-Boltzmann velocity distribution.

%
%
%
%
\subsection{Cavity -- Waveguide Coupling Model}
We model the coupling of the waveguides to the cavity using the coupled-mode equations of Haus \cite{haus1984} (see also \cite{vahala2004}). This method is related to the E\&M approach taken in \cite{jacobs2009all}, and we get nearly identical results using either technique.  We will label the ports as done in Fig. \ref{fig:eit_switch}.  The input port will be called ``in'' and corresponds to the upper left waveguide in the figures, and the two output ports will be labelled the ``drop'' port and ``through'' port for the lower left and upper right waveguides respectively.

The equation governing the field amplitude inside the resonator for a single input field is is given by
\begin{equation}
\label{eq:resonatoramplitude}
\frac{d a}{d t} = - i \Delta a - \frac{1}{2}(\kappa_0 + \kappa_e + \kappa_1 + \kappa_2) a + i \sqrt{\kappa_1}s_{in},
\end{equation}
where $a$ is the field amplitude inside the resonator, $\Delta=(\omega - \omega_0)$ is the detuning of the field with frequency $\omega$ from the cavity resonance of frequency $\omega_0$, $\kappa_0$ is the intrinsic cavity linewidth, $\kappa_e$ is related to the absorption rate of the atoms, $\kappa_1$ and $\kappa_2$ are the cavity -- waveguide coupling rates for waveguides 1 and 2 respectively, and $|s_{in}|^2$ is the input power. 

The steady state solution to Eq. \eqref{eq:resonatoramplitude} is
\begin{equation}
\label{eq:resonatorsteadystate}
a = i \frac{2\sqrt{\kappa_1}}{2i \Delta + (\kappa_0 + \kappa_e + \kappa_1 + \kappa_2)}s_{in}.
\end{equation}
The power out the through port is related to the power in plus the component that couples out from the resonator
\begin{equation}
\label{eq:powerthrough}
s_{t} = s_{in} + i \sqrt{\kappa_1}a = \frac{2i \Delta + (\kappa_0 + \kappa_e - \kappa_1 + \kappa_2)}{2i \Delta + (\kappa_0 + \kappa_e + \kappa_1 + \kappa_2)}s_{in}.
\end{equation} 
The signal power out the drop port is related to the power that couples out from the resonator,
\begin{equation}
\label{eq:powerdrop}
s_d = i \sqrt{\kappa_2}a = \frac{-2\sqrt{\kappa_1\kappa_2}}{2i \Delta + (\kappa_0 + \kappa_e + \kappa_1 + \kappa_2)}s_{in}.
\end{equation}
The through port and drop port transmission rates can now be easily calculated.  They are simply $T=|s_t|^2/|s_{in}|^2$ and $D=|s_d|^2/|s_{in}|^2$ which yield
\begin{subequations}
\label{eq:throughdroptransmission}
\begin{align}
T & = \frac{4\Delta^2 + (\kappa_0 + \kappa_e - \kappa_1 + \kappa_2)^2}{4\Delta^2 + (\kappa_0 + \kappa_e + \kappa_1 + \kappa_2)^2} \\
D & = \frac{4\kappa_1 \kappa_2}{4\Delta^2 + (\kappa_0 + \kappa_e + \kappa_1 + \kappa_2)^2}.
\end{align}
\end{subequations}

The intrinsic quality factor is defined as $Q_0=\omega_0/\kappa_0$ and the external quality factor is $Q_{ext}=\omega_0 / (\kappa_e+\kappa_1 + \kappa_2)$.  By controlling the atomic absorption rate $\kappa_e$ we can modify the external $Q$ of the atom-resonator system causing the coupling conditions to change, resulting in the ability to switch between the drop port and through port.

%
%
%
%
\subsection{Cavity Field -- Atomic Interaction Model}
The interaction of the evanescent fields to the Rubidium surrounding the cavity is the source of the added loss $\kappa_e$ in the waveguide--resonator coupling model above.  To give performance estimates for the switch, we calculate the average absorption coefficient of the signal beam in the resonator.

To calculate $\kappa_e$ we first solve the atomic equations, including Doppler broadening, given in \eqref{eq:densitymatrix} for a particular value of the signal and control beam fields.  The base absorption coefficient from standard quantum-optics theory is given by:
\begin{equation}
\label{eq:absorptioncoef}
\alpha(\Omega_s,\Omega_c) = \frac{4 N d^2 \omega \chi_{12}(\Omega_s,\Omega_c)}{\hbar \epsilon_0 c},
\end{equation}
where $\chi_{12}(\Omega_s,\Omega_c) = {\rm Im}[\rho_{12}(\Omega_s,\Omega_c)]/\Omega_{s}$ is related to the linear susceptibility and is dependent upon: signal and control beam Rabi frequencies, the density $N$ of Rubidium atoms, the dipole moment $d$ and the angular frequency $\omega$ of the 780 nm transition.  Here we explicitly note the functional dependence of all terms that depend upon the Rabi frequencies (field amplitudes) for clarity.

The small mode volume and high Q of the microdisk produce large intra-cavity intensities for even modest input power levels.  This can induce large splitting of the intermediate level.  However, the field distribution of the control beam outside the resonator falls off sharply as shown in Fig. \ref{fig:FieldProfile}.  To capture this, we compute the density matrix elements for each Rabi frequency corresponding to the different field values, and then average over the field distributions of both the signal and control fields as shown in Eq. \eqref{eq:aveabsorptioncoef} below.  These solutions are then averaged and weighted by the normalized signal beam intensity, to give the average absorption coefficient for the signal beam.  This average absorption coefficient $\bar{\alpha}$ is given by
\begin{equation}
\label{eq:aveabsorptioncoef}
\bar{\alpha} = \int_{V_e}\alpha(\Omega_s,\Omega_b)w(\vec{r}) dV,
\end{equation}
where the integral is taken to extend over the volume external to the resonator, denoted by $V_e$, and the weighting function
\begin{equation}
\label{eq:weightingfunction}
w(\vec{r}) = \frac{|\Omega_s(\vec{r})|^2}{\int_{V_t} |\Omega_s(\vec{r})|^2 dV}
\end{equation}
scales with the signal beam intensity.  

The normalization integral in the denominator of the weighting function extends over the entire volume of integration $V_t$, including the region interior to the resonator, as opposed to the integral in Eq. \eqref{eq:aveabsorptioncoef} which is only over the region exterior to the resonator.  This accounts for the fact that the field inside the resonator does not interact with the Rubidium thus reducing the overall strength of the interaction.  This weighting also accounts for the fact that the scattering cross section is larger near regions of high signal beam intensity than regions of low intensity.  We assume that both beams are contained in the same cavity mode for this calculation.

We can relate the average absorption coefficient, $\bar{\alpha}$ to the external cavity loss rate $\kappa_e$ through the simple relation $\kappa_e = c\bar{\alpha}$, where $c$ is the speed of light in the cavity.  Thus, by changing the atomic absorption coefficient $\bar{\alpha}$ through EIT, we can modify the total cavity Q, resulting in changing coupling conditions and switching.

\begin{figure}[h!]
\centering
\includegraphics[width=6.1cm]{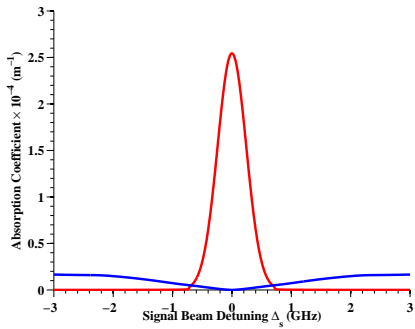}
\caption{Average absorption coefficient, denoted by $\bar{\alpha}$ in the text, of the signal beam with the control beam on (blue curve) and control beam off (red curve), plotted as a function of the signal beam detuning, $\Delta_s$. With the control beam on, the standard Doppler broadened line is split.  Because of the non-uniform control field intensity, the usual Autler--Townes splitting is replaced with the absorption profile shown in blue. }
\label{subFigA}
\end{figure}

\begin{figure}[h!]
\hspace{-0.3cm}
\subfigure[Through port transmission equal bandwidth]{
\centering
\includegraphics[width=4.3cm]{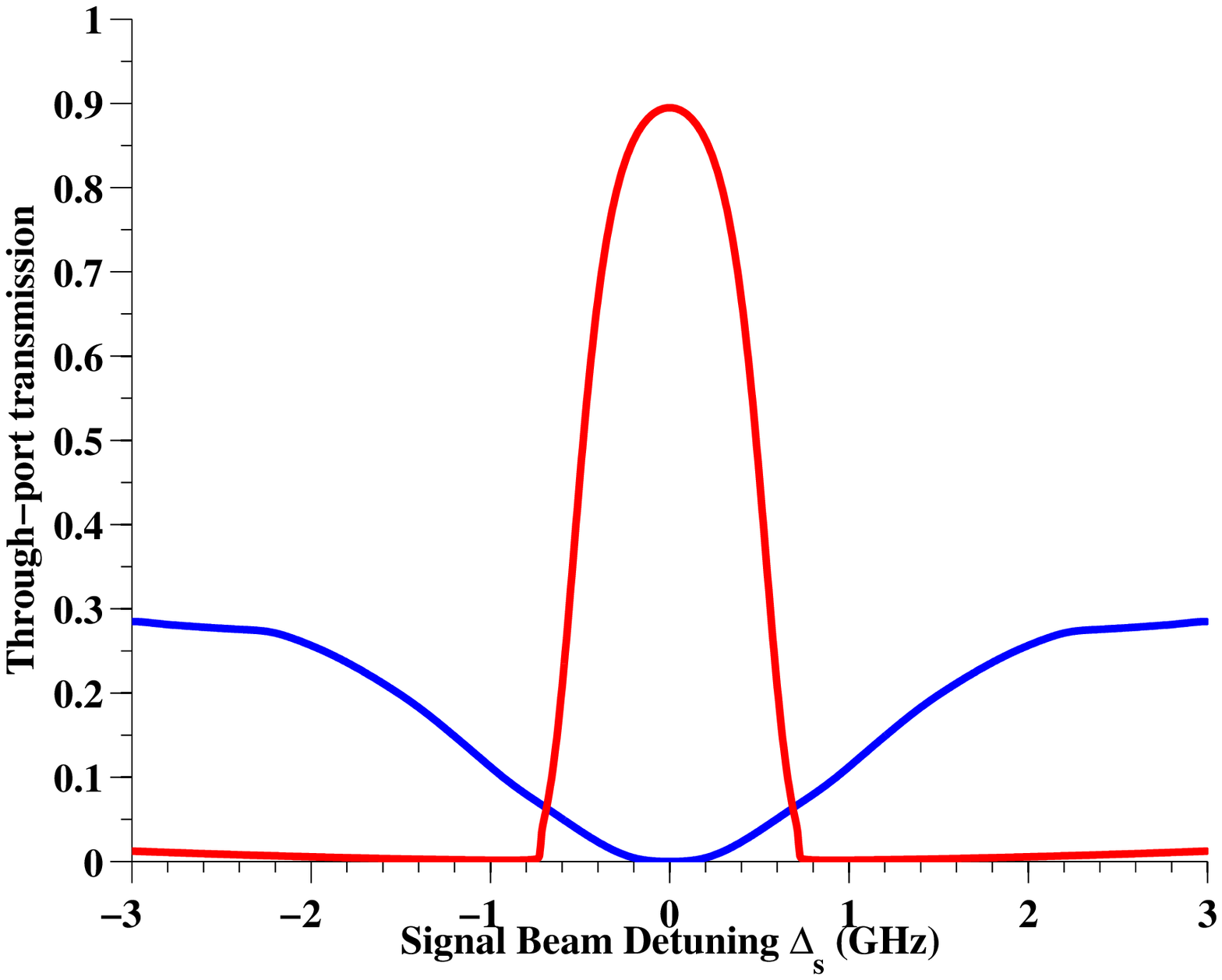}
\label{subFigB}
}
\hspace{-0.6cm}
\subfigure[Drop port transmission equal bandwidth]{
\centering
\includegraphics[width=4.3cm]{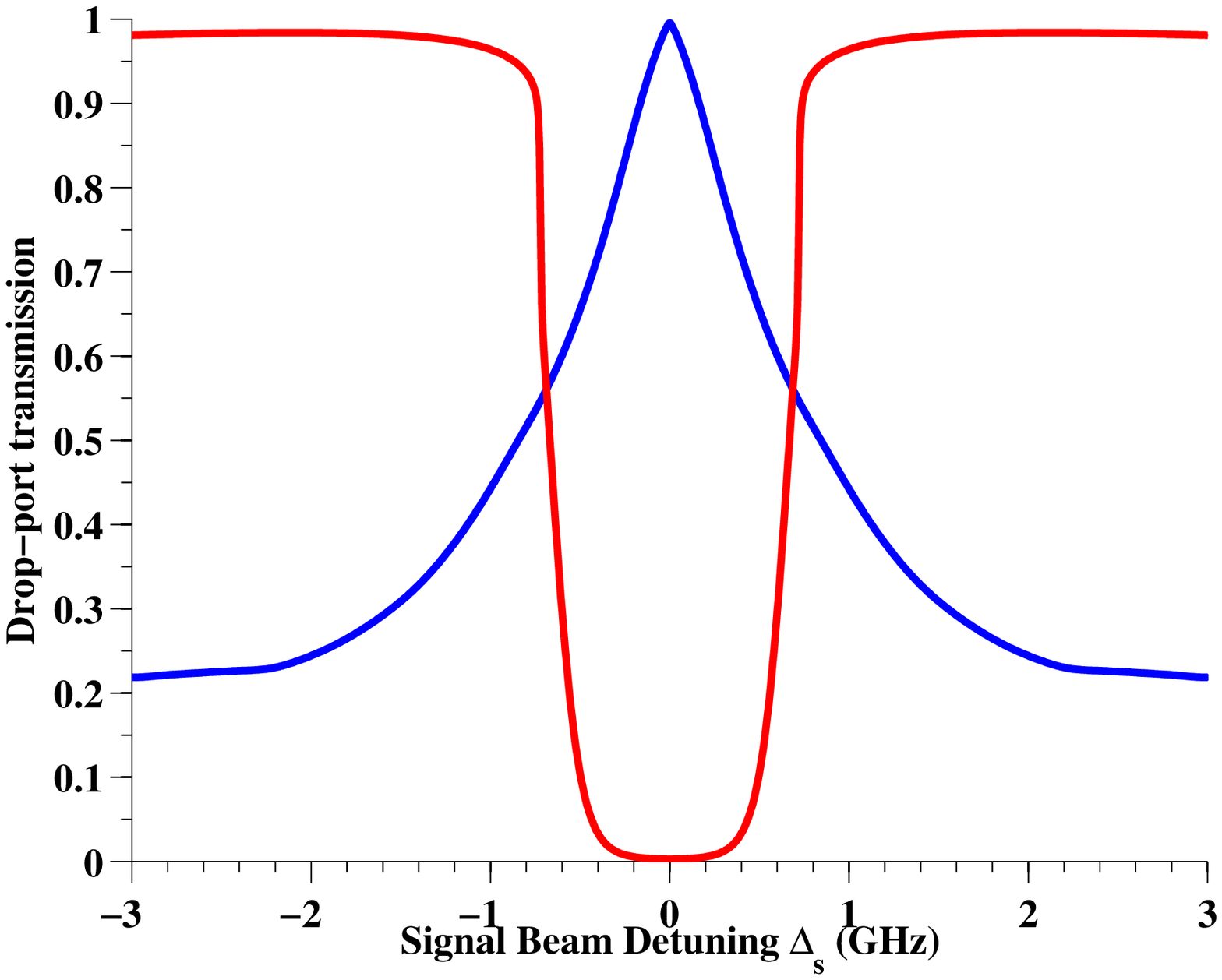}
\label{subFigC}
}
\subfigure[Through port transmission equal contrast]{
\hspace{-0.3cm}
\includegraphics[width=4.3cm]{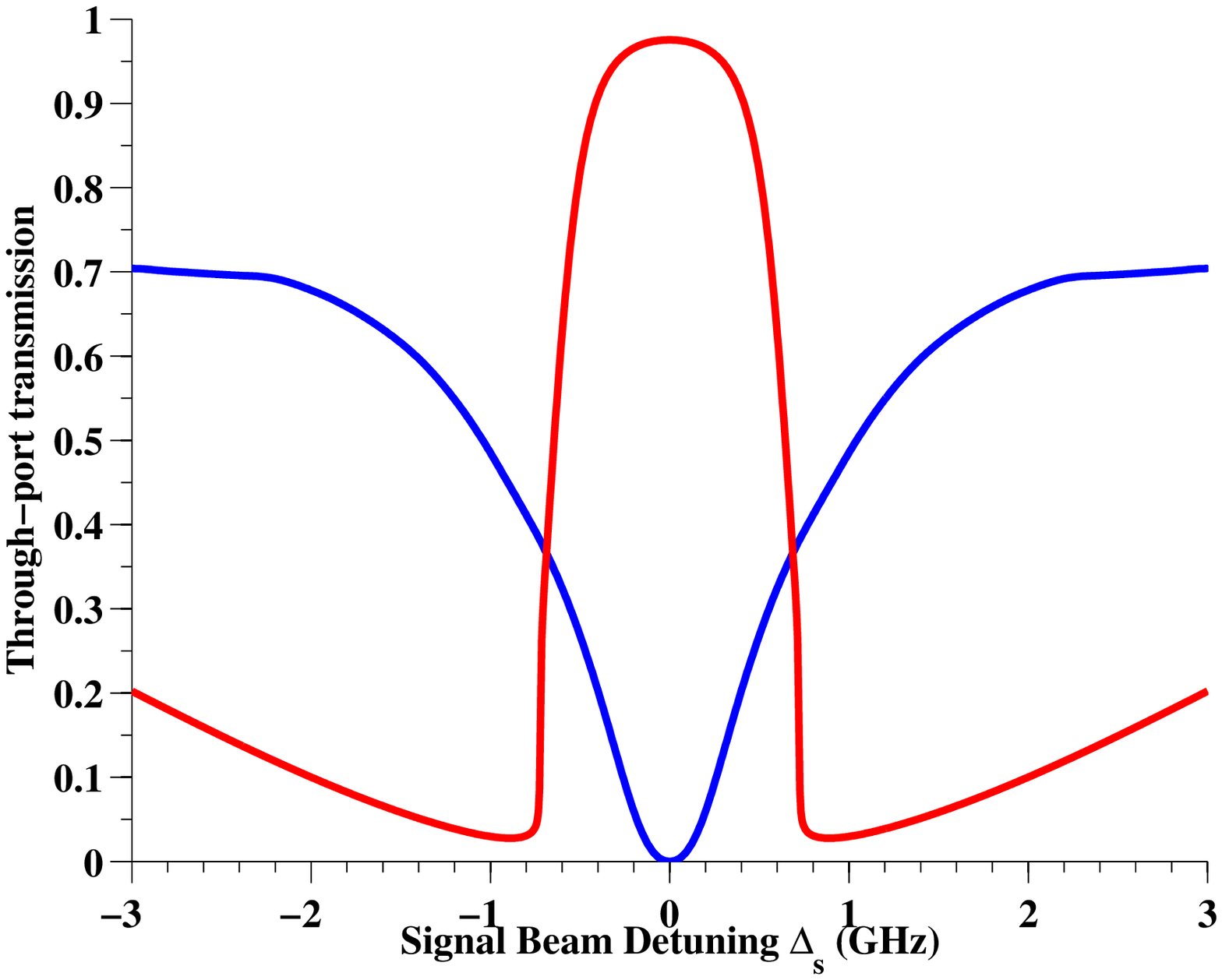}
\label{fig:equalContrastA}
}
\hspace{-0.53cm}
\subfigure[Drop port transmission equal contrast]{
\centering
\includegraphics[width=4.3cm]{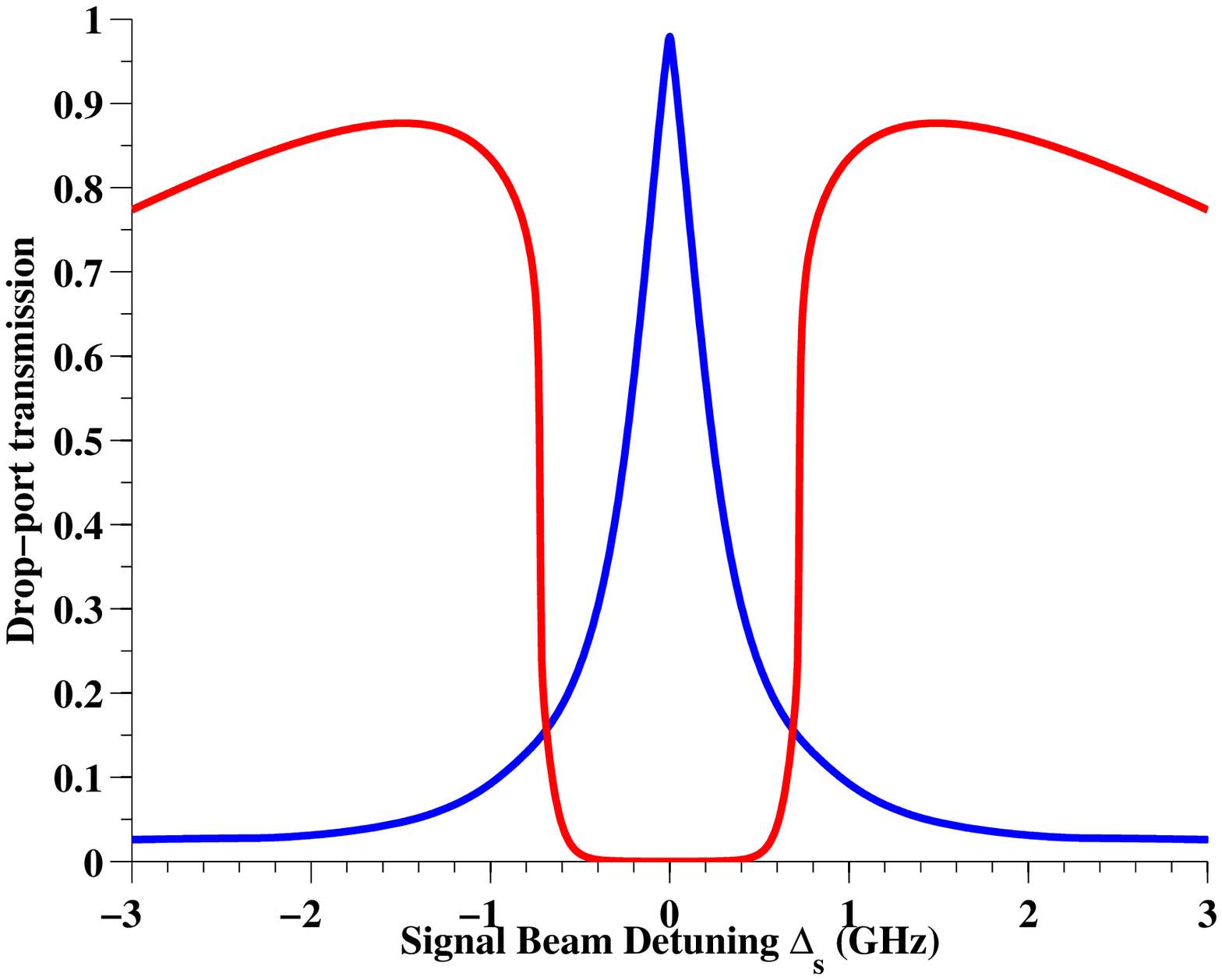}
\label{fig:equalContrastB}
}
\caption{\label{fig:TransmissionPlots} Through and drop port transmission plots for the parameters given in Tables \ref{table:results} and \ref{table:results2}.  Each plot provides the transmission percentage as a function of the signal laser detuning from the cavity resonance.  The red curve corresponds to the case where \ac{EIT} is off, and the blue curve corresponds to the case when the \ac{EIT} control beam is on.  In figures (a) and (b) we use the parameters in Table \ref{table:results} to equalize the bandwidth in both ports.  In figures (c) and (d) we use the parameters in Table \ref{table:results2} where the coupling condition was chosen such that the contrast was equalized.}
\end{figure}

\begin{center}
\begin{table}[h!]
\centering\caption{\label{table:parameters}Parameters used in atomic simulation.}
\begin{ruledtabular}
  \begin{tabular}{l  l  l  l }
    {\bf Variable} & {\bf Description} & {\bf Value} & {\bf Units} \\
    \hline
    $P_c$ & Control Beam Power & $2.0$ & $\mu$W \\
    $P_s$ & Signal Beam Power & $20$ & fW \\
    $A_m$ & Effective Mode Area & $8.2\times 10^{-10}$ & cm${}^2$\\
    $N$ & Rubidium Density & $5.0\times 10^{12}$ & cm${}^{-3}$ \\
    $d_s$ & Lower dipole moment & $2.1\times 10^{-29}$ & m$\cdot$C \\
    $d_c$ & Upper dipole moment & $4.6\times 10^{-30}$ & m$\cdot$C \\
    $\Gamma_{12}$ & Lower decay rate & $6.0$ & MHz \\
    $\Gamma_{23}$ & Upper decay rate & $280$ & kHz \\
    $\Gamma_{13}$ & Upper decay rate & $150$ & kHz \\
    $\sigma_D$ & Doppler linewidth & $240$ & MHz \\
  \end{tabular}
  \end{ruledtabular}
\end{table}
\end{center}

%
%
%
%
\section{Results}
Using the methodology above, the average absorption coefficient of the signal beam is calculated and plotted in Fig. \ref{subFigA}, with the control beam on and off (blue and red curves respectively).  All relevant atomic parameters for this simulation are given in Table \ref{table:parameters}.  The spectral shape of the absorption coefficient is modified from the typical double-Lorentzian peaks associated with Autler--Townes splitting because of Doppler broadening and the transverse variation of control beam intensity in the evanescent field.   This causes the signal beam to see a distribution of \ac{EIT} windows rather than a single split line as would be the case in a uniform field.  Very large Autler--Townes splitting on the order of GHz is possible, because the high-Q and small mode volume of the resonator can generate high-intensity fields with relatively modest input powers. 

Upon calculating the average absorption coefficient of the signal beam in the resonator, we estimate the transmission of the through-port (Fig. \ref{subFigB}) and drop-port (Fig. \ref{subFigC}) with and without the control beam, using the formulas in Eq. \eqref{eq:throughdroptransmission}.  We take the waveguides to be strongly over-coupled to the cavity with $\kappa_1$ and $\kappa_2$ roughly two orders of magnitude larger than $\kappa_0$.  The coupling rates were chosen to equalize the bandwidth on the through-port and drop-port.  With this constraint, we estimate the on-resonant switching contrast to be 50 dB in the through-port and 25 dB in the drop-port, along with only 0.5 dB and 0.02 dB of loss in each respectively.  We define the bandwidth as the point at which the switching contrast reaches 20 dB, on Figs. \ref{subFigB} and \ref{subFigC}.  This gives a bandwidth of approximately 516 MHz for each output port.  These results are tabulated in Table \ref{table:results}.

Switching performance trade-offs can be analyzed by varying the coupling rates between the waveguides and cavity.  In the previous example, we chose to equalize the bandwidth in the through-port and drop-port.  This required strongly over-coupling the waveguides to the cavity.  We can instead choose to equalize the contrast, and by extension the loss in each port.  To do this, we lower the fixed waveguide-cavity coupling rates which could be done by increasing the distance between the waveguides and the cavity for example.  These results are shown in Figs. \ref{fig:equalContrastA} and \ref{fig:equalContrastB}, with the corresponding performance metrics given in Table \ref{table:results2}.  On cavity resonance we estimate 38 dB switching contrast with only 0.1 dB loss in each port.  The tradeoff for equalized contrast, and lower loss is a reduced bandwidth in the through-port of 330 MHz, however the drop-port bandwidth is increased a comparable amount due to conservation of energy constraints.

\begin{table}
\centering\caption{\label{table:results}Switch performance results - equal bandwidth}
\begin{ruledtabular}
  \begin{tabular}{l l  l  l }
    {\bf Variable} & {\bf Description} & {\bf Value} & {\bf Units} \\
    \hline
    $Q_0$ & Intrinsic Quality Factor & $3.6\times 10^6$ & - \\
    $\kappa_1$ & Waveguide 1 coupling & 26.7 & GHz \\
    $\kappa_2$ & Waveguide 2 couplng & 26.7 & GHz \\
    \hline
    \multicolumn{4}{c}{{\bf Results}} \\
     & Through port loss & 0.5 & dB \\
     & Through port contrast  & 50 & dB \\
     & Drop port loss  & 0.02 & dB \\
     & Drop port contrast  & 25 & dB \\
     & Through port bandwidth & 500 & MHz \\
     & Drop port bandwidth & 500 & MHz \\
  \end{tabular}
  \end{ruledtabular}
  \end{table}
%
%
%
%
\section{Conclusions}
We have presented numerical simulations demonstrating the effectiveness of using \ac{EIT} together with the classical Zeno effect to create an all-optical switch in a micro-resonator.  Because of the large resonant build-up that the micro-resonators allow, even with modest input powers, very high intensity control fields can be created which can create excellent switching contrast with low loss.  In addition, this method is compatible with a very low input power signal.  These properties suggest that such a device could be suitable for switching single-photon intensities, ideal for some quantum information applications.  

  \begin{table}[h!]
  \centering\caption{\label{table:results2}Switch performance results - equal contrast}
\begin{ruledtabular}
  \begin{tabular}{l l  l  l }
    {\bf Variable} & {\bf Description} & {\bf Value} & {\bf Units} \\
    \hline
    $Q_0$ & Intrinsic Quality Factor & $3.6\times 10^6$ & - \\
    $\kappa_1$ & Waveguide 1 coupling & 5.9 & GHz \\
    $\kappa_2$ & Waveguide 2 couplng & 5.9 & GHz \\
    \hline
    \multicolumn{4}{c}{{\bf Results}} \\
     & Through port loss & 0.1 & dB \\
     & Through port contrast  & 38 & dB \\
     & Drop port loss  & 0.1 & dB \\
     & Drop port contrast  & 38 & dB \\
     & Through port bandwidth & 170 & MHz \\
     & Drop port bandwidth & 840 & MHz \\
  \end{tabular}
  \end{ruledtabular}
  \end{table}

\section*{Acknowledgments}
Funding was provided in part by IRAD and the DARPA ZOE program (Contract No. W31P4Q-09-C-0566).  We acknowledge thought provoking discussions with Jim Franson and Todd Pittman.

\end{document}